THE DISCORDANCE OF INDIVIDUAL RISK ESTIMATES AND THE REFERENCE

CLASS PROBLEM

Author:         Ralph H. Stern PhD,MD

e-mail:         stern@umich.edu

Address:        CVC Cardiovascular Medicine

                1500 East Medical Center Drive SPC5853

                Ann Arbor, MI 48109-5853

ABSTRACT

Multivariate methods that relate outcomes to risk factors have been adopted clinically to individualize treatment. This has promoted the belief that individuals have a "true" or unique risk.

The logic of assigning an individual a single risk value has been criticized since 1866. The reason is that any individual can be simultaneously considered a member of different groups, with each group having its own risk level (the reference class problem).

Lemeshow et al. provided well-documented examples of remarkable discordance between predictions for an individual by different valid predictive methods utilizing different risk factors. The prevalence of such discordance is unknown as it is rarely evaluated, but must be substantial due to the abundance of risk factors.

Lemeshow et al. cautioned against using ICU mortality predictions for the provision of care to individual patients. If individual risk estimates are used clinically, users should be aware that valid methods may give very different results.

A number of clinical tools are available to predict the future clinical course of patients.  Usually these are multivariate methods that predict clinical outcome based on the presence and/or magnitude of multiple risk factors.  It is commonly assumed that these methods can correctly assess an individual's risk.  However it is well-documented that valid predictive methods can produce very different risk assessments for the same individual.  If individuals possess "true" or unique risks, then the lack of agreement between predictive methods might be dismissed as statistical variation.  The alternative explanation, that individuals do not possess a "true" or unique risk that can be accurately assessed, has not been considered in the medical literature.   We review the discordance of individual risk estimates, show that there is not a unique way to distribute risk among individuals in a population, and discuss prior work from the literature of probability and philosophy criticizing the concept of individual risk.

Discordance of Individual Risk Assessments

Lemeshow et al. applied two models for ICU mortality (Apache II and MPMII at 24 hours) with similar discrimination and calibration to 11,320 patients and the predictions for each individual were compared.[1] A scatterplot of the paired predictions (Figure 1) showed remarkable discordance.  The mean difference in mortality probabilities assigned by the two models was only 0.01, but the standard deviation was

0.17. For 19.1% of patients, the difference between the mortality probabilities assigned by the two models was between 10 and 20%. For 19.8% of patients, the difference between the mortality probabilities assigned by the two models was greater than 20%. Similar results were obtained comparing another model, SAPS II, to Apache II and to MPMII at 24 hours. The authors concluded that the error rate does not support using models for deciding on the provision of care for individual patients, although they could be used for stratification for clinical trials, quality assessment of ICU performance, and hospital reimbursement, and discussions of prognosis.

The following examples are consistent with Lemeshow et al.'s description of poor agreement. They do not follow a standard protocol, so the conclusions reached by the authors are emphasized over detailed presentations of the comparisons.

Steyerberg et al. studied the agreement between 4 models of mortality after acute myocardial infarction (GUSTO-I, Belgium, TIMI-II, and GISSI-II) when applied to 40,830 patients included in the GUSTO-I trial.[2] Weighted kappa statistics ranged from 0.33 to 0.52 for the different model comparisons. They concluded that "Models with different predictors may have a similar validity while the agreement between predictions for individual patients is poor."

Pinna-Pintor et al. used 4 models for mortality after coronary artery bypass surgery in 418 patients.[3] The scores for the patients who died were on average higher than for those who survived, but were quite variable in some models. They concluded the models were "very inaccurate to predict mortality in individual patients."

Orr et al. used 4 models for mortality after coronary artery bypass surgery in 868 patients.[4] There was moderate correlation between scores for the same individual with the correlation coefficients ranging from 0.60 to 0.72.  They concluded that "The use of these models for individual patient risk estimations is risky because of the marked discrepancies in individual predictions created by each model."

Reynolds et al. evaluated the agreement between three models for cardiovascular risk on a simulated population of 10,000 individuals.[5] They used Venn diagrams to assess concordance for patients assigned to different risk strata and found that the three methods identify risk populations that differed significantly. They concluded that there was a lack of concordance between the models for identification of high risk individuals.

Other publications describing limited agreement of individual predictions for surgical and ICU patients include Iezzoni et al.,[6] Johnston,[7] and Moreno et al...[8]

McTiernan et al. compared two breast cancer risk estimates in 491 women with a family history of breast cancer and found only moderate (r=0.55) correlation.[9] They concluded that presenting both estimates to women being counseled might be the best option.

McGuigan et al. compared two breast cancer risk estimates in 111 women and found substantially different estimates of breast cancer risk for some patients.[10]

Multiple Equivalent Methods for Risk Stratification

The mathematical problem of risk stratification of a population does not have a unique solution, so disagreement between valid predictive methods should not be a surprise.  An example illustrates this point.  Assume a population of 6 patients with an event rate of 50%. Consider clinical predictive methods that accurately separate this population into high risk (66% event rate) and low risk (33% event rate) subpopulations of equal size. The 9 ways this could be achieved are illustrated in figure 2. Although these risk stratifications are equivalent, with identical calibration and discrimination, no individual is consistently assigned to the same risk group by all the methods.

Discussion

Validation and comparison of clinical predictive methods is usually limited to calibration and discrimination. The above examples were primarily located by citation analysis of the study of Lemeshow et al., but this important paper has been cited infrequently. It would be difficult to comprehensively review the literature as the phenomenon is unnamed and a wide range of methods have been utilized on the rare occasions when discordance of individual risk estimates has been evaluated.

The work of Lemeshow et al. clearly demonstrates that valid predictive methods may produce markedly different results for an individual. The other studies cited support this finding. When agreement is evaluated, the prediction tools compared should have comparable accuracy (calibration) and discrimination. This was the case in the study of Lemeshow et al., but not always the case in the other studies reviewed.

Although the issue of differing predictions for the same individual is rarely addressed in the medical literature, this has been discussed in abstract terms for many years in the literature on probability. Gaulte recently identified the relevance of this literature for clinical predictions.[11] John Venn wrote[12] in 1866: "It is obvious that every individual thing or event has an indefinite number of properties or attributes observable in it, and might therefore be considered as belonging to an indefinite number of different classes of things..." As these classes may have different risks, one individual may be assigned many different risks, a paradox known as the reference class problem.[13] von Mises referred to designating any particular value as an individual's risk as "total nonsense."[14]

Since there is not a unique way to risk stratify a population, disagreement between risk assessments for an individual must occur. Most clinical outcomes can be associated with multiple risk factors that are neither necessary nor sufficient for an event nor are that risky. Thus multiple panels of risk factors can be identified, each panel assigning an individual to a different reference class, producing discordant risk estimates. When panels include many of the same factors, e.g. when models differ by a single risk factor,[15] discordance will be minimized. When panels include few of the same risk factors, discordance will be maximized. Of the 12 risk factors in Apache II[16] and 13 risk factors in MPM,[17] the models in figure 1, only coma, creatinine, and $P_{O2}$ were shared. Panels that include the most risk factors or that include the most parsimonious set of risk factors are no more accurate than other panels (although they may be more discriminating) and should not be interpreted as providing "true" risks. In fact models of superior discrimination assign patients to a broader range of risks setting the stage for large absolute differences in individual risk estimates. In this setting the choice of panels can change a high risk patient to a low risk patient.

Individual risk estimates derived from models of association are conditional probabilities, dependent on the risk factors used. For this reason, it is more correct to refer to the probability of a myocardial infarction in 10 years given the Framingham risk factors than to the probability of a myocardial infarction in 10 years. It is true that a clinician's subjective judgment does not enter into the calculation of an individual risk assessment, but it does enter into the choice of the assessment method. Although risk factors may be inherent properties of an individual, risk is in the eye of the beholder. Rather than considering predictive methods to be pinpointing the risk of an individual, it is useful to consider them to be risk stratifying a population.

      Discordance of individual risk estimates does not weaken the economic rationale for their use in allocation of resources, but it does weaken the clinical rationale. Lemeshow[1] argued that ICU mortality estimates should not be used for individual patient care decisions. If individual risk estimates are used clinically, users should be aware that valid methods may give very different results.

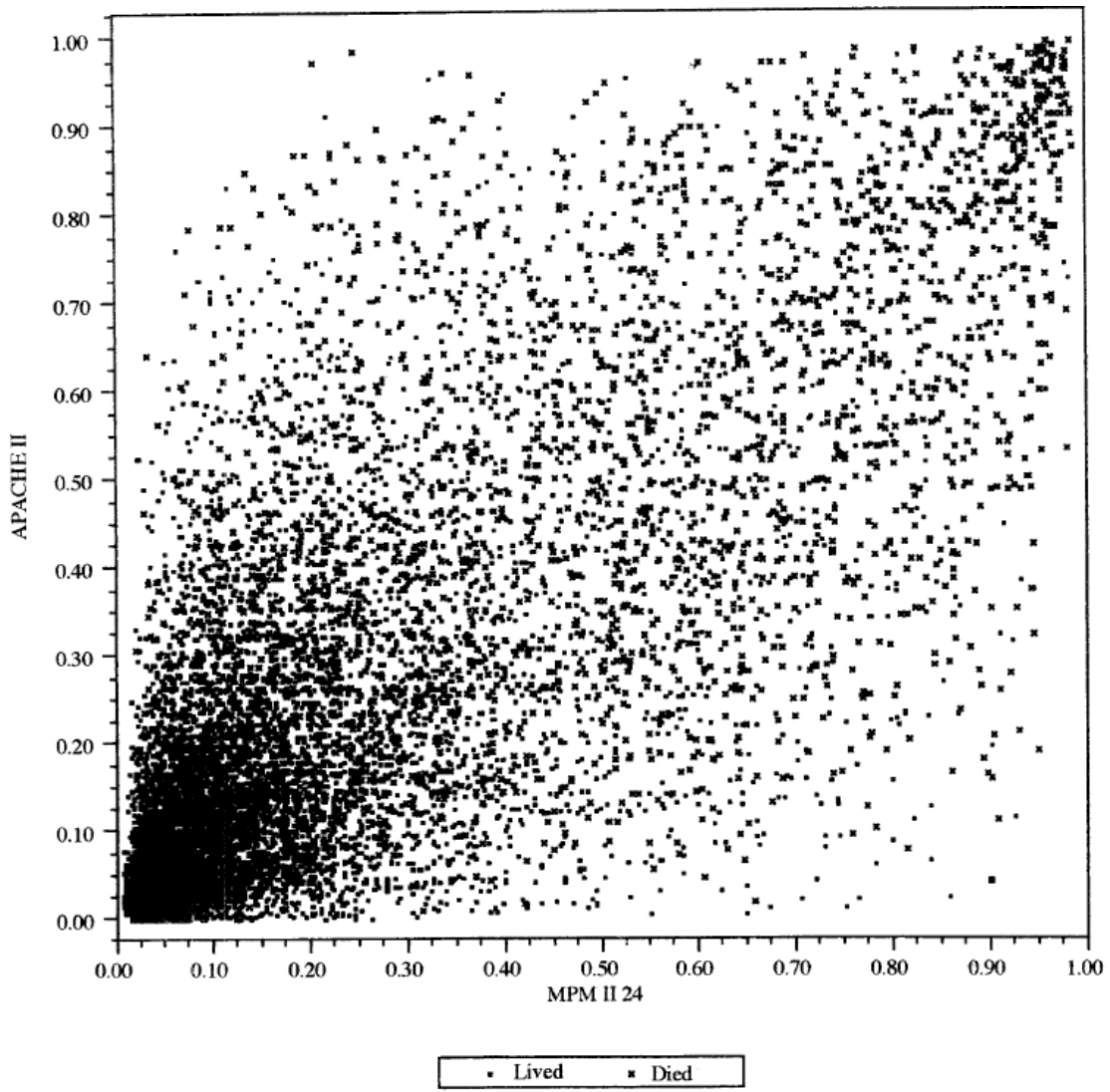

Figure 1. Scatterplot of Apache II probability versus MPM II$_{24}$ probability by vital status at hospital discharge (reproduced from Lemeshow et al.[1] with kind permission of Springer Science and Business Media)

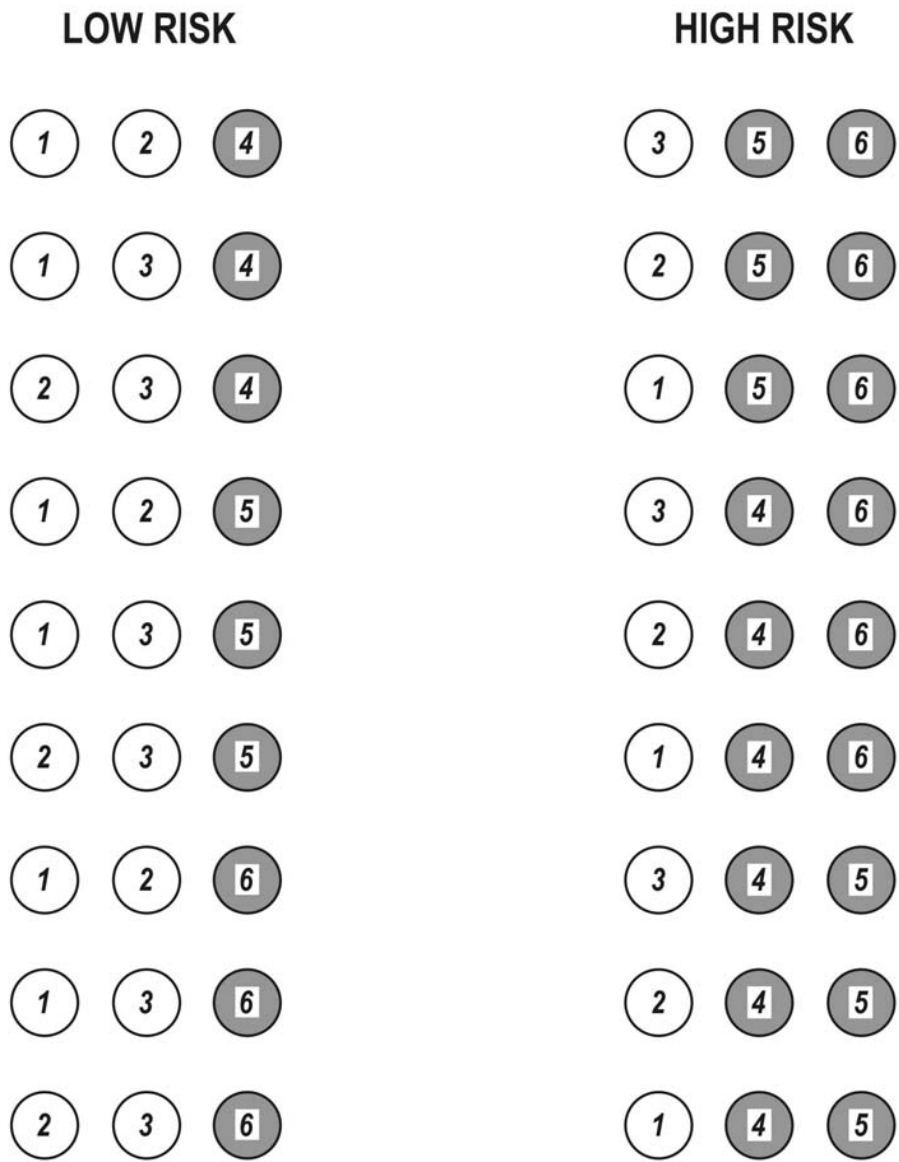

Figure 2. Example of Equivalent But Different Risk Stratifications

The three subjects without events (1, 2, 3) are depicted by white circles and the three subjects with events (4, 5, 6) are depicted by black circles. Each row represents a different risk stratification method for identifying high and low risk subpopulations.